\documentclass[amsmath,amssymb,pra]{revtex4}
\usepackage{hyperref}
\usepackage{makeidx,endnotes,fancyhdr}
\usepackage{showidx}
\usepackage{amsmath,amssymb,theorem,mathrsfs,epsf,eepic}
\usepackage{epsf,epsfig}

\newcommand{\ket}[1]{|#1\rangle}

\def\Tr{\operatorname{Tr}}

\def\>{\rangle}\def\<{\langle}
\def\sH{\mathscr{H}}
\def\sK{\mathscr{K}}

\def\geq{\geqslant}\def\leq{\leqslant}

\begin{document}

\title{Universal and phase covariant superbroadcasting for mixed qubit states}
\author{Francesco Buscemi}\email{buscemi@fisicavolta.unipv.it}
\author{Giacomo Mauro D'Ariano}\email{dariano@unipv.it} \author{Chiara
  Macchiavello}\email{chiara@unipv.it} \author{Paolo
  Perinotti}\email{perinotti@fisicavolta.unipv.it}\affiliation{QUIT
  Group, Dipartimento di Fisica ``A.  Volta'', Universit\`a di Pavia,
  via Bassi 6, I-27100 Pavia, Italy}\homepage{http://www.qubit.it/}
\date{\today}

\begin{abstract}
  We describe a general framework to study covariant symmetric
  broadcasting maps for mixed qubit states. We explicitly derive the
  optimal $N\to M$ superbroadcasting maps, achieving optimal
  purification of the single-site output copy, in both the universal
  and the phase covariant cases. We also study the bipartite
  entanglement properties of the superbroadcast states.
\end{abstract}

\maketitle

\section{Introduction}

``Information'' is by its nature {\em broadcastable}. What happens
when information is quantum and we need to distribute it among many
users?  Indeed, this may be useful in all situations where quantum
information is required in sharable form, e.~g. in distributed quantum
computation \cite{discom}, for quantum shared secrecy \cite{shacrip},
and, generally, in quantum game-theoretical contexts \cite{gam}.
However, contrarily to the case of classical information, which can be
distributed at will, broadcasting quantum information can be done only
in a limited fashion. Indeed, for pure states ideal broadcasting is
equivalent to the so-called ``quantum cloning'', which is impossible
due to the well-known ``no-cloning'' theorem
\cite{Wootters82,Dieks82,Yuen} (see also
\cite{noteclon,darianoyuen,herbert,peres}). The situation is more
involved when the input states are mixed, since broadcasting can be
achieved with an output joint state which is indistinguishable from
the tensor product of local mixed states from the point of view of
individual receivers. Therefore, the no cloning theorem cannot
logically exclude the possibility of ideal broadcasting for
sufficiently mixed states.

In Ref.~\cite{Barnum96} it was proved that perfect broadcasting is
impossible from a single input copy to two output copies for an input
set of non mutually commuting density operators.  This result was then
considered (see Refs. \cite{Barnum96} and \cite{clifton}) as an
evidence of the general impossibility of broadcasting mixed states
drawn from a non-commuting set in a more general scenario, where $N\geq 1$
equally prepared input copies are broadcast to $M>N$ users.  However,
for sufficiently many input copies $N$ and sufficiently mixed input
states the no-broadcasting theorem does not generally hold \cite{our},
and for input mixed states drawn from a noncommuting set it is
possible to generate $M>N$ output local mixed states which are
identical to the input ones, by a joint correlated state. Actually, as
proved in Ref.~\cite{our}, it is even possible to partially purify the
local state in the broadcasting process, for sufficiently mixed input
states.  Such a process of simultaneous purification and broadcasting
was named {\em superbroadcasting}.  For qubits, the fully covariant
superbroadcasting channel that maximizes the output purity (i. e. the
length of the output Bloch vectors of local states) when applied to
input pure states coincides with the optimal cloning
map~\cite{Werner98}.

The possibility of superbroadcasting does not increase the available
information about the original input state, due to unavoidable
detrimental correlations among the local broadcast copies, which do
not allow to exploit their statistics (a similar phenomenon was
already noticed in Ref.~\cite{Keyl01}). Essentially, superbroadcasting
transfers noise from local states to correlations.  From the point of
view of single users, however, the protocol is a purification in all
respects, and this opens new interesting perspectives in the ability
of distributing quantum information in a noisy environment, and
deserves to be analyzed in depth.  For qubits, it has been shown that
for universal superbroadcasting is possible with at least $N=4$ input
copies \cite{our}. Is this the absolute minimum number for
superbroadcasting, or does it hold only for this particular set of
input states?  In this paper we will show that, indeed, for equatorial
mixed states of qubits the minimum number is $N=3$.  However, for
smaller non-commuting sets of qubit states the possibility of
superbroadcasting with only $N=2$ input copies is still an open
problem (for larger dimension $d>2$ it is possible to superbroadcast
also for $N=2$, see e.g. Ref. \cite{cvsb}).

We want to point put that clearly there are limitations to
superbroadcasting. The input state must be indeed sufficiently mixed,
since pure states cannot be broadcast by the no cloning theorem.
However, states with a pretty high purity can sill be superbroadcast,
e.~g.  for universally covariant superbroadcasting \cite{our} from
$N=4$ to $M=5$ it is possible to superbroadcast states with a Bloch
vector length up to 0.787 (0.935 for the phase covariant case). One
can achieve superbroadcasting with even higher input purity for
increasing $N$, approaching unit input Bloch vector length in the
limit of infinitely many input copies. There are also some limitations
in the absolute number of output copies for which one can achieve
superbroadcasting.  E.~g. in the universal case, for $N=4$ input
copies one can superbroadcast up to $M=7$ output copies, for $N=5$ up
to $M=22$, and for $N>5$ up to infinitely many.  The output purity is
clearly decreasing versus $M$.  In this paper we will further analyze
all limitations to superbroadcasting also for the case of equatorial
input qubits.

Regarding the possibility of achieving superbroadcasting
experimentally, the first route to explore is to use the same
techniques as for purification \cite{Cirac99}, since the
superbroadcasting map for $M>N$ generalizes the purification map using
the same protocol \cite{altro}. This transformation involves a
measurement of the total angular momentum of the qubits, then an
optimal Werner cloning~\cite{Werner98} in the universal case (or an
optimal phase covariant cloning~\cite{purequbitqutrit} in the phase
covariant case). Another possibility is to use the methods of
Ref.~\cite{Buscemi03} in order to classify all possible unitary
realizations, and then seek for experimentally achievable ones using
current technology.

In this paper we present phase covariant superbroadcasting, and also
give a complete derivation of the universal superbroadcasting map,
presented in Ref.~\cite{our}. The two maps are derived in a unified
theoretical framework. The paper is organized as follows. In Section
\ref{s:symm-broad} we introduce some preliminary notions regarding
symmetric covariant maps. In Sects. \ref{s:univ} and \ref{s:pc} we
give a complete derivation of the optimal broadcasting maps in the
universal case and in the phase covariant case respectively.  In Sect.
\ref{sec:concurrence} we study the entanglement properties of the
states of two copies at the output of the universal and the phase
covariant broadcasting maps. Finally, in Sect. \ref{s:conc} we
summarize and comment the main results of this paper.  At the end of
the paper we report the details of the calculations needed to derive
the results presented in three appendices.

\section{Symmetric qubits broadcasting}\label{s:symm-broad}

In this Section we introduce in a unified theoretical framework some
preliminary concepts that will be employed to describe covariant
symmetric qubit broadcasting maps. These concepts will be then
specified to the universal and the phase covariant cases in the subsequent
sections.  A main tool we will extensively use in deriving the optimal
maps is the formalism of the Choi-Jamio\l kowski isomorphism
\cite{choi-jam} between completely positive (CP) maps $\mathcal E$
from states on the Hilbert space $\sH$ to states on the Hilbert space
$\sK$, and positive bipartite operators $R$ on $\sK\otimes\sH$. Such
an isomorphism can be specified as follows
\begin{equation}\label{eq:choi-jam}
R\doteq\mathcal{E\otimes
    I}(|\Omega\>\<\Omega|)\longleftrightarrow\mathcal{E}(\rho)\doteq\Tr_\sH[\openone\otimes\rho^T\ R],
\end{equation}
where $\Omega$ is the non normalized maximally entangled state
$\sum_m|\psi_m\>\otimes|\psi_m\>$ in $\sH\otimes\sH$, $\mathcal{I}$
gives the identity transformation and $X^T$ denotes the transposition
of the operator $X$ on the same basis $|\psi_m\>$ used in the
definition of $\Omega$.

In terms of $R$, the trace-preservation condition for the map
$\mathcal E$ reads
\begin{equation}\label{trace-pres-abovo}
\Tr_\sK[R]=\openone_\sH\,.
\end{equation}
Suppose that the map $\mathcal E$ is covariant under the action of a
group $\bf G$. In this case the covariance property is reflected to
the form of the operator $R$ by the following correspondence
\begin{equation}\label{eq3}
  \mathcal{E}(U_g\rho U_g^\dag)\equiv V_g\mathcal{E}(\rho)V_g^\dag\Longleftrightarrow[V_g\otimes U_g^*,R]=0.
\end{equation}
In the above expression $U_g$ and $V_g$ are the unitary
representations of ${\bf G}\ni g$ on the input and output spaces
respectively, while $X^*$ denotes complex conjugation on the fixed
basis $|\psi_m\>$.  In this framework it is also possible to study
group-invariance properties of the map $\mathcal E$ in terms of the
operator $R$.  In this case we have the following equivalences
\begin{equation}\label{invariance1}
  \mathcal{E}(U_g\rho U_g^\dag)\equiv \mathcal{E}(\rho)\Longleftrightarrow[\openone\otimes U_g^*,R]=0,
\end{equation}
and
\begin{equation}\label{invariance2}
  \mathcal{E}(\rho)\equiv V_g\mathcal{E}(\rho)V_g^\dag\Longleftrightarrow[V_g\otimes\openone,R]=0.
\end{equation}
The above expressions refer to invariance properties on the input and
output spaces respectively.

In the following we will consider maps $\mathcal B$ from states of $N$
qubits to states of $M$ qubits, namely CP maps from states on
$\sH=(\mathbb{C}^2)^{\otimes N}$ to states on
$\sK=(\mathbb{C}^2)^{\otimes M}$. We will consider in particular
symmetric broadcasting maps, namely transformations where all
receivers get the same reduced state. The figures of merit which are
commonly used are invariant under permutations of the output copies,
and this allows to assume that the output state of a broadcasting map
is permutation invariant without loss of generality. Moreover, since
the input consists in $N$ copies of the same state, there is no loss
of generality in requiring that the map is also invariant under
permutations of the input copies. These two properties, according to
Eqs. (\ref{invariance1}, \ref{invariance2}), can be recast as follows
\begin{equation}
[\Pi_\sigma^M\otimes\Pi_\tau^N,R]=0,\qquad\forall\sigma,\tau,
\label{perminv}
\end{equation}
where $\Pi_\sigma^M$ and $\Pi_\tau^N$ are representations of the
output and input copies permutations, respectively. Notice that
permutations representations are all real, hence
$\Pi_\sigma^*=\Pi_\sigma$.

A useful tool to deal with unitary group representations $U_g$ of a
group $\bf G$ on a Hilbert space $\sH$ is the Wedderburn decomposition
of $\sH$
\begin{equation}
\sH\simeq\bigoplus_\mu\sH_\mu\otimes\mathbb{C}^{d_\mu}\,,
\end{equation}
where the index $\mu$ labels the equivalence classes of irreducible
representations which appear in the decomposition of $U_g$. The spaces
$\sH_\mu$ support the irreducible representations and
$\mathbb{C}^{d_\mu}$ are the multiplicity spaces, with dimension
$d_\mu$ equal to the degeneracy of the $\mu$-th irreducible
representation. Correspondingly the representation $U_g$ decomposes as
\begin{equation}\label{eq:wedder-for-U}
  U_g=\bigoplus_\mu U_g^\mu\otimes\openone_{d_\mu},
\end{equation}
where $\openone_{d_\mu}$ is shorthand for $\openone_{{\mathbb
    C}^{d_\mu}}$. By Schur's Lemma, every operator $X$ commuting with
the representation $U_g$ in turn decomposes as
\begin{equation}
  X=\bigoplus_\mu\openone_{\sH_\mu}\otimes X_{d_\mu}.
\end{equation}

In the case of permutation invariance, the so-called Schur-Weyl
duality \cite{fulton} holds, namely the spaces $\mathbb{C}^{d_\mu}$
for permutations of $M$ qubits coincide with the spaces $\sH_\mu$ for
the representation $U_g^{\otimes M}$ of $\mathbb{SU}(2)$, where $U_g$
is the defining representation. In other words, a permutation
invariant operator $Y$ can act non trivially only on the spaces
$\sH_\mu$, namely it can be decomposed as
\begin{equation}\label{eq:schur-Weyl-dual-form}
Y=\bigoplus_\mu Y_\mu\otimes\openone_{d_\mu}.
\end{equation}
The Clebsch-Gordan series for the defining representation of
$\mathbb{SU}(2)$ is well-known in the literature
(see for example \cite{fulton,edmonds}), 
and its Wedderburn decomposition is given by
\begin{equation}\label{eq:wedderburn-for-su2}
  \sH\simeq\bigoplus_{j=j_0}^{M/2}\sH_j\otimes\mathbb{C}^{d_j}\,,
\end{equation}
where $\sH_j=\mathbb{C}^{2j+1}$, $j_0$ equals 0 for $M$ even, 1/2 for
$M$ odd, and
\begin{equation}
d_j=\frac{2j+1}{M/2+j+1}\binom{M}{M/2-j}.
\end{equation}\par

In the case of the broadcasting maps the Hilbert space $\sK\otimes\sH$
on which the operator $R$ acts, supports the two permutation
representations corresponding to the output and input qubits
permutations. Therefore it can be decomposed as
\begin{equation}
\sK\otimes\sH\simeq\left(\bigoplus_{j=j_0}^{M/2}\sH_j\otimes\mathbb{C}^{d_j}\right)\otimes\left(\bigoplus_{l=l_0}^{N/2}\sH_l\otimes\mathbb{C}^{d_l}\right)\,.
\end{equation}
By rearranging the factors in the above expression, we can recast the
decomposition in a more suitable way, namely
\begin{equation}
\sK\otimes\sH\simeq\bigoplus_{j=j_0}^{M/2}\bigoplus_{l=l_0}^{N/2}\left(\sH_j\otimes\sH_l\right)\otimes\left(\mathbb{C}^{d_j}\otimes\mathbb{C}^{d_l}\right)\,.
\label{shuf}
\end{equation}
The operator $R$, in order to satisfy the permutation invariance property 
\eqref{perminv}, 
according to Eq.~(\ref{eq:schur-Weyl-dual-form}),
can be written in the following form 
\begin{equation}
R=\bigoplus_{j=j_0}^{M/2}\bigoplus_{l=l_0}^{N/2} R_{jl}\otimes\left(\openone_{d_j}\otimes \openone_{d_l}\right)\,,
\label{symmbro}
\end{equation}
where the operators $R_{jl}$ act on $\sH_j\otimes\sH_l$.
Moreover, in order to fulfill the requirements of trace preservation and 
complete positivity, the operators $R_{jl}$ must satisfy the constraints
\begin{equation}
  R_{jl}\geq0\,,\quad\Tr_j[R_{jl}]=\frac{\openone_{2l+1}}{d_j},
\label{cpt}
\end{equation}
where $\Tr_j$ denotes the partial trace performed over the space
$\sH_j$ in the $j$-th term of the decomposition in Eq.~\eqref{shuf},
and $\openone_{2l+1}$ is shorthand for $\openone_{\sH_l}$.  We have
now all the tools to study symmetric qubits broadcasting devices.  In
this work we are interested in the case of covariant broadcasting
maps, which in general have to fulfill also the following covariance
condition under the representations $V_g^{\otimes N}$ and
$V_g^{\otimes M}$ of a group $\bf G$ (see Eq.~(\ref{eq3}))
\begin{equation}
\left[V_g^{\otimes M}\otimes {V_g^*}^{\otimes N},R\right]=0.
\end{equation}
The above condition gives a further constraint on the form of the
operators $R_{jl}$ in Eq. (\ref{symmbro}).  Actually, the group $V_g$
is in general just a subgroup of the defining representation $U_g$ of
$\mathbb{SU}(2)$, and therefore the representation $V_g^{\otimes
  M}\otimes {V_g^*}^{\otimes N}$ acts non trivially only on the
subspaces $\sH_j\otimes\sH_l$, which are the ones supporting the
operators $R_{jl}$.  In the next sections we will consider two
interesting cases, namely $V_g\equiv U_g$ and $V_g\equiv
V_\phi=e^{i\frac\phi2\sigma_z}$, corresponding to universal and
phase covariant broadcasting respectively, and we will see how the
form of the operators $R_{jl}$ depends on the particular choice of the
considered covariance group.

In addition to the Wedderburn decomposition and the related Schur-Weyl
duality reviewed above, another useful tool we will extensively use in
the following is a convenient decomposition of an $N$-partite state of
the form $\rho^{\otimes N}$, representing $N$ qubits all prepared in
the same generic state $\rho$. In Appendix \ref{appendixaa} we report
the complete derivation of the following identity, which was
originally presented in \cite{Cirac99},
\begin{equation}\label{eq:rho-decomposition}
  \rho^{\otimes N}=(r_+r_-)^{N/2}\bigoplus_{j=j_0}^{N/2}
\sum_{m=-j}^j\left(\frac{r_+}{r_-}\right)^m|jm\>\<jm|\otimes \openone_{d_j}.
\end{equation}
For the sake of simplicity, in the above expression we considered
density operators of the form $\rho=(\openone+r\sigma_z)/2$,
$r_\pm=(1\pm r)/2$, namely qubit states whose Bloch vector (of length
$r$) is aligned along the $z$ axis, and consequently the states
$|jm\>$ are eigenstates of the operator $J_z$ in the $j$
representation, namely $J_z^{(j)}=\sum_{m=-j}^jm|jm\>\<jm|$. Notice
actually that the \emph{total} angular momentum component $J_z$ of $N$
qubits is clearly permutation invariant and therefore it can be
written as
\begin{equation}
  J_z=\bigoplus_{j=j_0}^{N/2}J_z^{(j)}\otimes\openone_{d_j}.
\label{Jz}
\end{equation}
We want to point out that the
decomposition~(\ref{eq:rho-decomposition}) holds for any direction of
the Bloch vector, provided that the eigenstates of $J_z$ in Eq.
(\ref{eq:rho-decomposition}) are replaced by the eigenvectors of the
angular momentum component along the direction of the Bloch vector in
the single qubit state $\rho$.

We will prove in Appendix C that the single-site output copy
$\rho^\prime$ of a covariant broadcasting map commutes with the input
density operator $\rho$. In order to quantify the performance of the
broadcasting map $\mathcal B$ and to judge the quality of the
single-site output density operator $\rho'=\Tr_{M-1}[\mathcal
B(\rho^{\otimes N})]$ we will then evaluate the length $r^\prime$ of
its Bloch vector, namely
\begin{equation}
\Tr[\sigma_z\rho^\prime]=r^\prime\,.
\label{singsitra}
\end{equation}
Notice, moreover, that the length of a Bloch vector $r$ is simply
related to the purity $\Tr[\rho^2]$ of the density operator $\rho$ as
$\Tr[\rho^2]=(1+r^2)/2$.  Therefore, maximizing the output Bloch
vector length $r'$ is equivalent to maximizing the output single-site
purity.  Notice also that so far we cannot exclude that the input and
output Bloch vectors $r$ and $r'$ are antiparallel, and this just
implies that $r^\prime$ can range from $-1$ to $1$.

We will now show how to evaluate $r'$ according to
Eq.~(\ref{singsitra}), which is the main quantity that we will
consider in the next sections in the particular cases of universal and
phase covariant broadcasting.  The trace in Eq.~\eqref{singsitra} can
be computed by considering that the global output state
$\Sigma=\mathcal B(\rho^{\otimes N})$ of the $M$ copies is by
construction invariant under permutations, hence
\begin{equation}\label{eq:rprime-valutaz}
r^\prime=\Tr[\sigma_z\rho^\prime]=\Tr\left[\left(\sigma_z\otimes\openone^{\otimes M-1}\right)\Sigma\right]=\frac1{M!}\Tr\left[\sum_\sigma\Pi_\sigma\left(\sigma_z\otimes\openone^{\otimes M-1}\right)\Pi_\sigma \Sigma\right]\,.
\end{equation}
The last term on the r.h.s. of Eq. (\ref{eq:rprime-valutaz}) contains
a sum over the $M!$ possible permutations of the $M$ output qubits.
Notice that
\begin{equation}
  \frac1{M!}\sum_\sigma\Pi_\sigma(\sigma_z\otimes\openone^{\otimes (M-1)})\Pi_\sigma=\frac1M\sum_{i=1}^M\sigma_z^{(i)}=\frac2M J_z\,,
\end{equation}
where the operator $\sigma_z^{(i)}$ acts as $\sigma_z$ on the $i$-th
qubit and identically on the remaining qubits. Now, by exploiting the
permutation invariance of $\Sigma$, we can write
\begin{equation}
\Sigma=\bigoplus_{j=j_0}^{M/2}\Sigma_j\otimes\openone_{d_j}\,,
\end{equation}
and clearly
\begin{equation}\label{eq:rprime}
r^\prime=\frac 2M\sum_{j=j_0}^{M/2}d_j\Tr[J_z^{(j)}\Sigma_j]\,.
\end{equation}
The explicit expression of $r'$ in the universal case will be derived
in Sect. \ref{s:univ}. In the phase covariant case, we will see that
it is more convenient to take $\rho$ diagonal on the $\sigma_x$
eigenstates. The above formula in this case is just substituted by
\begin{equation}\label{eq:rprime-phase}
r^\prime=\frac 2M\sum_{j=j_0}^{M/2}d_j\Tr[J_x^{(j)}\Sigma_j]\,
\end{equation}
and will be explicitly calculated in Sect. \ref{s:pc}.

We want to stress that maximization of the figure of merit $r'$ allows
to optimize the fidelity criterion as well. In fact, for $r'<r$ the
two criteria coincide, whereas for $r'\geq r$ one can always achieve
unit fidelity by suitably mixing the output state with optimal $r'$
and the maximally mixed one. On the other hand, direct maximization of
fidelity is not analytically feasible, since fidelity is a concave
function over the convex set of covariant maps, whence it is not
maximized by extremal maps.

Finally, we want to mention that in the next sections we explicitly
maximize the scaling factor for $N$ inputs and $M$ outputs
$p^{N,M}(r)\equiv r'/r$, which can be referred to as {\em shrinking
  factor} or {\em stretching factor}, depending whether it is smaller
or greater than 1, respectively. It is obvious that this maximization
is equivalent to maximizing $r'$. Superbroadcasting corresponds to the
cases where $p(r)>1$.


\section{universal case}\label{s:univ}
In this Section we will give the explicit derivation of the optimal
universal broadcasting maps. Starting from the general broadcasting
map described in the previous section we have to impose in this case
the additional constraint
\begin{equation}\label{univ1}
[U_g^{\otimes M}\otimes {U_g^*}^{\otimes N},R]=0\,,
\end{equation}
where $U_g$ is the defining representation of the group
$\mathbb{SU}(2)$. For the defining representation $U_g$ the following
property holds
\begin{equation}
U_g^*=\sigma_y U_g\sigma_y\,.
\end{equation}
By exploiting such a property, the commutation relation (\ref{univ1})
can be written more conveniently as follows
\begin{equation}
[U_g^{\otimes (M+N)},S]=0\,,
\label{universales}
\end{equation}
where $S=(\openone^{\otimes M}\otimes\sigma_y^{\otimes
  N})R(\openone^{\otimes M}\otimes\sigma_y^{\otimes
  N})=\left(\openone^{\otimes M}\otimes e^{i\pi
    J_y}\right)R\left(\openone^{\otimes M}\otimes e^{-i\pi
    J_y}\right)$. The complete positivity and trace-preservation
constraints in terms of the operator $S$ are then equivalent to
\begin{equation}
S\geq0\,,\quad \Tr_{\sK}[S]=\openone_\sH\,.
\end{equation}
Upon defining $S_{jl}=\left(\openone_{2j+1}\otimes e^{i\pi
    J^{(l)}_y}\right)R_{jl}\left(\openone_{2j+1}\otimes e^{-i\pi
    J^{(l)}_y}\right)$, the constraints for complete positivity and
trace preservation are now given by the following conditions on the
operators $S_{jl}$
\begin{equation}
S_{jl}\geq0\,,\quad\Tr_j[S_{jl}]=\frac{\openone_{2l+1}}{d_j}.
\end{equation}
By exploiting the fact that the Clebsch-Gordan series for 
$\sH_j\otimes\sH_l$ is
just $\bigoplus_{J=|j-l|}^{j+l}\sH_J$, we can write
\begin{equation}
  \sK\otimes\sH=\bigoplus_{j=j_0}^{M/2}\bigoplus_{l=l_0}^{N/2}\bigoplus_{J=|j-l|}^{j+l}\sH^{j,l}_J\otimes\mathbb{C}^{d_j}\otimes\mathbb{C}^{d_l}\,.
\label{parwed}
\end{equation}
Notice that this is not the Wedderburn decomposition, since not all
the subspaces $\sH^{j,l}_J\simeq\mathbb{C}^{2J+1}$ support
inequivalent representations. However, the Wedderburn decomposition
can be recovered by a suitable rearrangement that takes into account
the repetitions of the same representation $J$. Using the
decomposition \eqref{parwed} we can formulate the constraint
\eqref{universales} in terms of the operators $S_{jl}$ as follows
\begin{equation}\label{onlyonerunning}
S_{jl}=\bigoplus_{J=|j-l|}^{j+l}s_{j,l}^J P_{j,l}^{J}\,,
\end{equation}
where, by complete positivity, the coefficients $s_{j,l}^J$ are real
and positive, and $P_{j,l}^J$ is the projection of the space
$\sH_j\otimes\sH_l$ onto the $J$ representation, satisfying
\begin{equation}
\Tr_j\left[P^J_{j,l}\right]=\frac{2J+1}{2l+1}\openone_{2l+1},\qquad\Tr_l\left[P^J_{j,l}\right]=\frac{2J+1}{2j+1}\openone_{2j+1}.
\end{equation}
The set of projectors $P_{j,l}^J$ is clearly orthogonal. The
trace-preservation constraint (\ref{trace-pres-abovo}) can now be
written as
\begin{equation}
  \bigoplus_{l=l_0}^{N/2}\sum_{j=j_0}^{M/2}\sum_{J=|j-l|}^{j+l}s_{j,l}^J\frac{2J+1}{2l+1}d_j\left(\openone_{2l+1}\otimes\openone_{d_l}\right)=\openone_\sH,
\end{equation}
which is equivalent to the conditions
\begin{equation}\label{eq:polytope}
\sum_{j=j_0}^{M/2}\sum_{J=|j-l|}^{j+l}s_{j,l}^J\frac{2J+1}{2l+1}d_j=1\,,
\quad\forall l\,.
\end{equation}
Along with the complete positivity constraint $s_{j,l}^J\geq0$, Eq.
(\ref{eq:polytope}) defines a convex polyhedron whose extremal points
are classified by functions $j=j_l$ and $J=J_l$
\begin{equation}
  s_{j,l}^{J}=\frac{2l+1}{2J_l+1}\frac1{d_{j_l}}\delta_{j,j_l}\delta_{J,J_l}.
\end{equation}
The classification of symmetric universally covariant maps is then
completely determined in terms of the vectors $j_l$ and $J_l$, whose
elements can range from $j_0$ to $M/2$ and from $|j_l-l|$ to $j_l+l$,
respectively. Extremal maps then correspond to the following form for
the operators $S$
\begin{equation}
  S=\bigoplus_{l=l_0}^{N/2}\frac{2l+1}{2J_l+1}\frac1{d_{j_l}}P^{J_l}_{j_l,l}\otimes\openone_{d_{j_l}}\otimes\openone_{d_l}.
\label{S}
\end{equation}
The optimization of the figure of merit $r^\prime$ or, equivalently,
of the scaling factor $p(r)$ can be obtained by explicit calculation
from Eq.~(\ref{eq:rprime}). The output state $\Sigma$ of the
broadcasting map applied to an input state $\rho^{\otimes N}$ can be
represented as
\begin{equation}
  \Sigma=\Tr_\sH[\openone^{\otimes M}\otimes(\sigma_y\rho^T\sigma_y)^{\otimes N}\ S]=\Tr_\sH[\openone^{\otimes M}\otimes{\tilde\rho}^{\otimes N}\ S]\,,
\label{maponinp}
\end{equation}
where $\tilde\rho$ denotes the orthogonal complement of $\rho$, which
just corresponds to the change $r\to-r$ (or, equivalently, $r_\pm\to
r_\mp$). Using the decomposition in Eq.~(\ref{eq:rho-decomposition})
for ${\tilde\rho}^{\otimes N}$ and the form~(\ref{S}) for the operator
$S$, we can express Eq.~\eqref{maponinp} as follows
\begin{equation}
  \Sigma=(r_+r_-)^{N/2}\sum_{l=l_0}^{N/2}\frac{2l+1}{2J_l+1}\frac{d_l}{d_{j_l}}\sum_{n=-l}^l\left(\frac{r_-}{r_+}\right)^{n}\Tr_l\left[\left(\openone_{2j_l+1}\otimes|ln\>\<ln|\right)\ P^{J_l}_{j_l,l}\right]\otimes\openone_{d_{j_l}}.
\end{equation}

We can now use Eq.~(\ref{eq:rprime}) to evaluate the scaling factor,
namely
\begin{equation}\label{eq:to-calculate}
p^{N,M}(r)=\frac2{Mr}(r_+r_-)^{N/2}\sum_{l=l_0}^{N/2}\frac{2l+1}{2J_l+1}d_l\sum_{n=-l}^l\left(\frac{r_-}{r_+}\right)^{n}\Tr\left[J_z^{(j_l)}\otimes|ln\>\<ln|\ P^{J_l}_{j_l,l}\right].
\end{equation}
In Appendix \ref{appendixb} we report the explicit calculation of
$p^{N,M}(r)$ and we show that it can be written in the following form
\begin{equation}\label{eq:univ-pr-to-opt}
  p^{N,M}(r)=\frac2{Mr}(r_+r_-)^{N/2}\sum_{l=l_0}^{N/2}\beta(J_l,j_l,l)d_l\sum_{n=-l}^l n\left(\frac{r_-}{r_+}\right)^{n}\,,
\end{equation}
where
\begin{equation}
  \beta(J,j,l)=\frac{J(J+1)-j(j+1)-l(l+1)}{2l(l+1)}\,.
\label{beta}
\end{equation}
Since $r_-\leq r_+$, the sum $\sum_{n=-l}^l n (r_-/r_+)^n$ in Eq.
(\ref{eq:univ-pr-to-opt}) is always negative. Therefore, the function
$p^{N,M}(r)$ is maximized by the choice of $J_l$ and $j_l$ minimizing
$\beta$, which clearly implies $J_l=|j_l-l|$. The form of the
coefficient $\beta(J_l,j_l,l)$ for $j_l<l$ is given by
\begin{equation}
\beta(l-j_l,j_l,l)=-\frac{j_l}l\,,
\label{beta1}
\end{equation}
whereas for $j_l\geq l$ we have 
\begin{equation}
\beta(j_l-l,j_l,l)=-\frac{j_l+1}{l+1}\,.
\label{beta2}
\end{equation}
In both cases $\beta$ is minimized by choosing the maximum value of
$j_l$, and therefore the maximum scaling factor is achieved by
$j_l=M/2$. For $M>N$ the optimal value of the figure of merit is then
univocally determined by the value of the function $\beta$
\begin{equation}
\beta(M/2-l,M/2,l)=-\frac{M+2}{2(l+1)}\,,
\end{equation}
while the optimal scaling factor is given by
\begin{equation}
p^{N,M}(r)=-\frac{M+2}{Mr}(r_+r_-)^{N/2}\sum_{l=l_0}^{N/2}\frac{d_l}{l+1}\sum_{n=-l}^l n\left(\frac{r_-}{r_+}\right)^{n}.
\end{equation}
The corresponding output state takes the form
\begin{equation}\label{eq:universal-global-output-state}
\begin{split}
  \Sigma=&(r_+r_-)^{N/2}\sum_{l=l_0}^{N/2}
  \frac{2l+1}{M-2l+1}d_l\times\\
  &\sum_{n=-l}^l\sum_{m=-M/2}^{M/2}
  \left\langle\frac{M}{2}m,ln\right|\left.\frac{M}{2}-l,m+n\right\rangle^2
  \left(\frac{r_-}{r_+}\right)^{n}\left|\frac M2,m\right\rangle\left\langle\frac M2,m\right|,
\end{split}
\end{equation}
where $\langle \frac{M}{2}m,ln|\frac{M}{2}-l,m+n\rangle$ denote the 
Clebsch-Gordan coefficients.

\begin{figure}
\includegraphics[width=9cm]{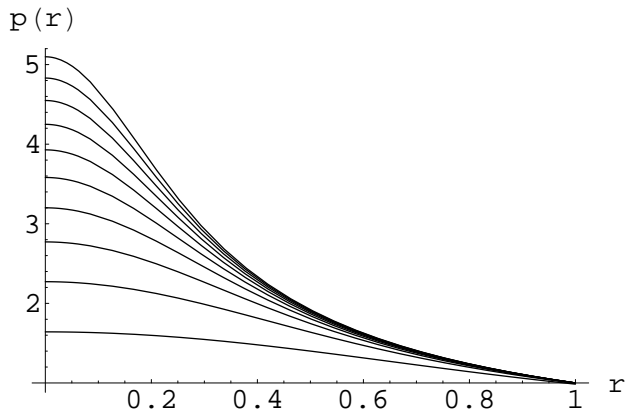}
\caption{Optimal scaling factor $p^{N,N+1}(r)=r^\prime/r$ versus $r$
  for the universal broadcasting, for $M=N+1$ and $N$ ranging from 10
  to 100 in steps of 10. Notice that there is a wide range of values
  of $r$ such that $p^{N,N+1}(r)>1$, corresponding to
  superbroadcasting.}\label{prplotz}
\end{figure}

As mentioned in the previous section, in Appendix \ref{appendixa} we
prove that the single-site reduced output state $\Tr_{M-1}[\Sigma]$
commutes with $\sigma_z$, hence $p^{N,M}(r)$ is definitely a scaling
factor.  Two interesting cases we will consider in the following are
the ones with $M=N+1$ and $M=\infty$, for which the scaling factor
takes the explicit forms
\begin{equation}
\begin{split}
  &p^{N,N+1}(r)=-\frac{N+3}{(N+1)r}(r_+r_-)^{N/2}\sum_{l=l_0}^{N/2}\frac{d_l}{l+1}\sum_{n=-l}^l n \left(\frac{r_-}{r_+}\right)^{n}\,,\\
  &p^{N,\infty}(r)=-\frac1r(r_+r_-)^{N/2}\sum_{l=l_0}^{N/2}\frac{d_l}{l+1}\sum_{n=-l}^l
  n\left(\frac{r_-}{r_+}\right)^{n}\,.
\end{split}
\end{equation}
The function $p^{N,N+1}(r)$ is plotted in Fig. \ref{prplotz} for $N$
ranging from 10 to 100 in steps of 10. We can see that for a suitable
range of values of $r$ the scaling factor is larger than one. This
corresponds to a broadcasting process with an increased single-site
purity at the output with respect to the input. This phenomenon occurs
for $N\geq4$. In this case $p(r)$ is actually a stretching factor, and
we call such a phenomenon superbroadcasting. The maximum value of $r$
such that it is possible to achieve superbroadcasting will be referred
to as $r_*(N,M)$ and it is a solution of the equation
\begin{equation}
  p^{N,M}(r_*)=1.
\end{equation}

\begin{figure}
  \includegraphics[width=9cm]{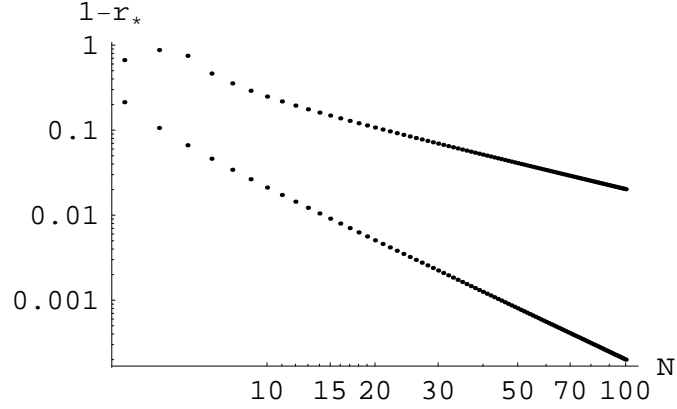}
  \caption{The logarithmic plot reports the behaviour in the universal
    case of $1-r_*(N,N+1)$ and $1-r_*(N,M_*(N))$ for $4\leq N\leq100$.
    The upper line, corresponding to $M=M_*(N)$, has a power law
    $1/N$.  The lower line, corresponding to $M=N+1$, has a power law
    $2/N^2$.}\label{rstar}
\end{figure}

It is clear that the optimal scaling factor for fixed $N$ is a non
increasing function of $M$. Actually, by contradiction, suppose that
the map with $M+K$ output copies has a higher purity than the optimal
map with $M$ copies. Then one could trace over $K$ copies from the
former map, and he would obtain a map with $M$ output copies with
purity higher than the optimal, which is obviously absurd. This
implies that in general $r_*(N,M)<r_*(N,M+K)$, and for large values of
$K$ superbroadcasting may not be possible anymore. The maximum $M$
such that superbroadcasting can be achieved for $N$ input copies will
be referred to as $M_*(N)$. It turns out that, apart from the values
$N=4,5$ for which we have $M_*(4)=7$ and $M_*(5)=21$, for $N\geq6$ one
has $M_*(N)=\infty$, namely superbroadcasting is possible for any
number of output copies. In Fig.~\ref{rstar} we report the values of
$1-r_*(N,N+1)$ and $1-r_*(N,M_*(N))$ for $4\leq N\leq100$. By a
numerical analysis we have evaluated the power laws for the two
curves, which turn out to be in good agreement with $2/N^2$ and $1/N$,
respectively.

We want to point out that for input pure states ($r=1$) only the term
with $l=N/2$ in the expression (\ref{S}) is significant. The optimal
map then corresponds to the optimal universal cloning for pure states
derived in \cite{Werner98}.

\section{phase covariant case}\label{s:pc}
In this section we study the case of symmetric phase covariant
broadcasting, where we restrict our attention to input states lying on
an equator of the Bloch sphere, say the $xy$-plane. The equatorial
qubit density operator in this case has the explicit form
$\rho=(\openone+r\cos\phi\sigma_x+r\sin\phi \sigma_y)/2$.  The
starting point, as in the case of universal broadcasting, is the
requirement of permutation invariance (\ref{perminv}) for input and
output copies, that leads to the form (\ref{symmbro}) of the operator
$R$.  Moreover, in this case we demand covariance under the action of the
group of rotations along the $z$-axis $V_\phi=e^{i\phi\sigma_z/2}$.
We proceed analogously to the case of universal broadcasting.  By
imposing the covariance condition for the map we require invariance of
the $R$ operator, namely $[V_\phi^{\otimes M}\otimes V_\phi^{*\otimes
  N},R]=0$. By exploiting the Wedderburn decomposition
(\ref{eq:wedder-for-U}) for the operator $V_\phi^{\otimes M}$, namely
\begin{equation}
V_\phi^{\otimes M}=\bigoplus_{j=j_0}^{M/2}e^{i\phi J_z^{(j)}}
\otimes\openone_{d_j},
\end{equation}
the phase covariance requirement corresponds to the following
additional condition for the operators $R_{jl}$
\begin{equation}\label{eq:Rjl-covariance}
  [R_{jl},e^{i\phi J_z^{(j)}}\otimes
  e^{-i\phi J_z^{(l)}}]=0,\qquad\forall j,l,
\end{equation}
where $J_z^{(j)}$ is defined according to Eq.~(\ref{Jz}). A
convenient form for the operators $R_{jl}$ satisfying Eq.
(\ref{eq:Rjl-covariance}) is the following
\begin{equation}\label{eq:Rjl1}
R_{jl}=\sum_{n=-l}^{l}\sum_{n'=-l}^{l}\sum_{k=l-j}^{j-l}r_{n,n',k}^{jl}|j,n+k\>\<j,n'+k|\otimes|l,n\>\<l,n'|
\end{equation}
when $j\geq l$, and
\begin{equation}\label{eq:Rjl2}
R_{jl}=\sum_{m=-j}^{j}\sum_{m'=-j}^{j}\sum_{k=j-l}^{l-j}r_{m,m',k}^{jl}|j,m\>\<j,m'|\otimes|l,m+k\>\<l,m'+k|
\end{equation}
when $j<l$. Notice that there are two more running indices with
respect to the universal case (\ref{onlyonerunning}). The index $n'$
in Eq.~(\ref{eq:Rjl1}) simply allows for off-diagonal contributions
in the operator $R_{jl}$, while we will see that the index $k$, which
labels equivalence classes, is related to the direction of the reduced
output state Bloch vector.  In particular we will show that, in order
to get an equatorial output, the operators $R_{jl}$ have to be
symmetric in $k$, in the sense that
$r_{n,n',k}^{jl}=r_{n,n',-k}^{jl}$. Notice also that $k$ takes integer
values when $M-N$ is even and half integer values when $M-N$ is odd.

The trace-preservation condition (\ref{cpt}) now reads
\begin{equation}\label{phase-cpt}
\sum_{j=j_0}^{M/2}\sum_{k=-|l-j|}^{+|l-j|}
r_{n,n,k}^{jl}d_j=1,\qquad\forall l,n,
\end{equation}
and, analogously to the universal case, the fact that the operators
$R_{jl}$ are diagonal with respect to the indices $j$'s and $k$'s
implies that the extremal points are classified by functions
\begin{equation}
j=j_l,\qquad k=k_l,
\end{equation}
and satisfy
\begin{equation}\label{eq-mancante}
  r^{j_l,l}_{n,n,k_l}=\frac{1}{d_{j_l}},\qquad\forall l,n.
\end{equation}

We will now compute the output density operator and the scaling factor
for $N\to M$ phase covariant broadcasting maps.  Without loss of
generality, let us now consider an input state $\rho$ oriented along
the $x$-axis, namely $\rho=(\openone+r\sigma_x)/2$.  The density
operator $\rho^{\otimes N}$ can then be decomposed, analogously to Eq.
(\ref{eq:rho-decomposition}), as
\begin{equation}
  \rho^{\otimes N}=(r_+r_-)^{N/2}\bigoplus_{l=l_0}^{N/2}\sum_{n=-l}^l
\left(\frac{r_+}{r_-}\right)^n|l^x,n\>\<l^x,n|\otimes\openone_{d_l},
\label{decpc}
\end{equation}
where $|l^x,n\>$ is the eigenvector of $J_x^{(l)}$ corresponding to
the eigenvalue $n$. In the following, eigenvectors without explicit
specification of the superscript axis, such as $|jm\>$, are intended
to be along the $z$-axis, namely $|j^z,m\>$.  According to
Eq.~(\ref{eq:choi-jam}), the density operator $\Sigma$ on
$\sK\equiv(\mathbb{C}^{2})^{\otimes M}$, describing the output state
of the $M$ copies, can be written as
\begin{equation}\label{eq:form-for-Sigma}
\begin{split}
  \Sigma&=\Tr_\sH\left[\left(\openone^{\otimes M}\otimes(\rho^T)^{\otimes N}\right)R\right]=\Tr_\sH\left[\left(\openone^{\otimes M}\otimes\rho^{\otimes N}\right)R\right]\\
  &=(r_+r_-)^{N/2}\bigoplus_{j=j_0}^{M/2}\sum_{l,k}d_l\sum_{n,n'}\sum_{n''}r^{j,l}_{n,n',k}\left(\frac{r_+}{r_-}\right)^{n''}(W_l^\dag)_{n'',n}(W_l)_{n',n''}|j,n+k\>\<j,n'+k|\otimes\openone_{d_j},
\end{split}
\end{equation}
where $(W_k)_{ab}\equiv\<k,a|k^x,b\>$ are the entries of the Wigner
rotation matrix in the $k$ representation which rotates the
$z$-components into the $x$-components---in the usual notation (the
one that is found, for example, in \cite{edmonds}) such entries are
denoted as $d^{(k)}_{ab}\left(\beta\equiv\frac{\pi}{2}\right)$. As
discussed previously in Sect.~\ref{s:symm-broad}, the projection $r'$
along the $x$ axis (\ref{eq:rprime-phase}) of the Bloch vector of the
single-site output state is a convex (linear) function on the convex
set of phase covariant broadcasting maps, and therefore it achieves
its maximum on extremal broadcasting maps. Let the functions $j=j_l$
and $k=k_l$ denote an extremal map. Hence, starting from
Eq.~(\ref{eq:rprime-phase}) and specializing
Eq.~(\ref{eq:phase-form-pr}), derived in Appendix B, to the extremal
case $j=j_l$ and $k=k_l$, we can express the scaling factor in the
following form
\begin{equation}\label{rprime-to-opt}
  p^{N,M}(r)=\frac{4}{Mr}(r_+r_-)^{N/2}\sum_{l=l_0}^{N/2}d_{j_l}d_l
  \sum_{n=-l}^lr^{j_l,l}_{n,n+1,k_l}\left[\exp\left(J_x^{(l)}\log\frac{1+r}{1-r}
    \right)\right]_{n,n+1}\left[J_x^{(j_l)}\right]_{n+k_l,n+k_l+1},
\end{equation}
where $[X^{(j)}]_{n,m}$ denotes the matrix element of the operator
$X^{(j)}$ evaluated with respect to the eigenstates of $J_z^{(j)}$,
i.~e. $|jm\>$. The final form in Eq.~(\ref{rprime-to-opt}) is now
suitable to be optimized. First of all, since the matrix elements
$\left[J_x^{(j)}\right]_{n,m}$ are non-negative, the maximum purity is
reached by maximizing the off-diagonal elements of $R_{jl}$, namely
for rank-one $R_{jl}$ with all the matrix elements equal to
$1/d_{j_l}$ (see Eq.~(\ref{eq-mancante})). We now want to identify the
values of $j_l$ and $k_l$ corresponding to the optimal scaling factor
of the map.  The matrix elements of $J_x^{(j_l)}$ take the explicit
form
\begin{equation}\label{Jxme}
  \left[J_x^{(j_l)}\right]_{n+k_l,n+k_l+1}=\frac{1}{2}\sqrt{j_l(j_l+1)-(n+k_l)
    (n+k_l+1)}\;.
\end{equation}
Since the above matrix elements are maximized in the central block of
the matrix, the optimal map is achieved by choosing $k_l$ as close as
possible to zero, for all the values of $l$.  When $M-N$ is even, the
optimal choice corresponds to $k_l=0$ for all $l$.  When $M-N$ is odd
there are two equivalent possible choices for $k_l$, namely $k_l=\pm
1/2$, for each value of $l$.  Moreover, we set $j_l=M/2$ for all $l$,
namely as large as possible.  For $M-N$ even, the global output
$\Sigma$ and the scaling factor are given by
\begin{equation}\label{eq:phase-pr1}
\begin{split}
  &\Sigma=(r_+r_-)^{N/2}\sum_{l=l_0}^{N/2}d_l\sum_{n,n'=-l}^l\left[\exp\left(J_x^{(l)}\log\frac{1+r}{1-r}\right)\right]_{n,n'}\left|\frac{M}{2},n\right\>\left\<\frac{M}{2},n'\right|,\\
  &p^{N,M}_e(r)=\frac{4}{Mr}(r_+r_-)^{N/2}\sum_{l=l_0}^{N/2}d_l\sum_{n=-l}^{l-1}\left[\exp\left(J_x^{(l)}\log\frac{1+r}{1-r}\right)\right]_{n,n+1}\left[J_x^{(M/2)}\right]_{n,n+1}.
\end{split}
\end{equation}
For $M-N$ odd we have many more solutions, corresponding to all the possible 
combinations of $k_l=\pm 1/2$ for all values of $l$. As will be clear in the 
following discussion, we will examine the two cases of $k_l=1/2$ and
$k_l=-1/2$ for all $l$. In the former case we can write
\begin{equation}\label{eq:phase-pr2}
\begin{split}
  &\Sigma=(r_+r_-)^{N/2}\sum_{l=l_0}^{N/2}d_l\sum_{n,n'=-l}^{l}
  \left[\exp\left(J_x^{(l)}\log\frac{1+r}{1-r}\right)\right]_{n,n'}
  \left|\frac M2,n+\frac 12\right\>\left\<\frac M2,n'+\frac 12\right|,\\
  &p^{N,M}_o(r)=\frac{4}{Mr}(r_+r_-)^{N/2}\sum_{l=l_0}^{N/2}d_l\sum_{n=-l}^{l-1}\left[\exp\left(J_x^{(l)}\log\frac{1+r}{1-r}\right)\right]_{n,n+1}\left[J_x^{(M/2)}\right]_{n+1/2,n+3/2},
\end{split}
\end{equation}
while for $k_l=-1/2$ we have
\begin{equation}\label{eq:phase-pr3}
\begin{split}
  &\Sigma=(r_+r_-)^{N/2}\sum_{l=l_0}^{N/2}d_l\sum_{n,n'=-l}^{l}
  \left[\exp\left(J_x^{(l)}\log\frac{1+r}{1-r}\right)\right]_{n,n'}
  \left|\frac M2,n-\frac 12\right\>\left\<\frac M2,n'-\frac 12\right|,\\
  &p^{N,M}_o(r)=\frac{4}{Mr}(r_+r_-)^{N/2}\sum_{l=l_0}^{N/2}d_l
  \sum_{n=-l}^{l-1}\left[\exp\left(J_x^{(l)}\log\frac{1+r}{1-r}\right)
  \right]_{n,n+1}\left[J_x^{(M/2)}\right]_{n-1/2,n+1/2}.
\end{split}
\end{equation}
Notice that, since $\left[J_x^{(j)}\right]_{m,m+1}=
\left[J_x^{(j)}\right]_{-(m+1),-m}$ and the same property holds for
the matrix elements of any power of $J_x$, the scaling factors
corresponding to the extremal maps with $k_l=1/2$ and $k_l=-1/2$ are
exactly the same. This means that the Bloch vector components in the
$xy$-plane are scaled in the same way by the two maps.

We want to point out that an extremal map with $k_l\neq 0$ generates
output density operators with a non vanishing component of the Bloch
vector along the $z$ direction. Actually, for the input state
(\ref{decpc}) the output single-site density operator is given by
\begin{equation}\label{eq_che_manca}
\rho'=\frac{1}{2}(\openone+r'\sigma_x+\alpha_k \sigma_z)\;,
\end{equation}
where
\begin{equation}
\alpha_k=\sum_l(2l+1)\frac{d_l}{2^N}\frac{k_l}{M}\;.
\end{equation}
Optimal broadcasting maps, where the Bloch vector is just scaled along
its input direction, can then be obtained for odd values of $M-N$ by
equally mixing the two maps considered above, corresponding to
$k_l=1/2$ and $k_l=-1/2$.  As mentioned earlier, since the two maps
give the same scaling factor, their mixture does not compromise
optimality.  Notice that the optimal broadcasting maps we have derived
in this way are independent of the input state. In the limit of pure
input states, the above maps coincide with the optimal phase covariant
cloning for pure equatorial states presented in
Ref.~\cite{purequbitqutrit}.

\begin{figure}
  \includegraphics[width=9cm]{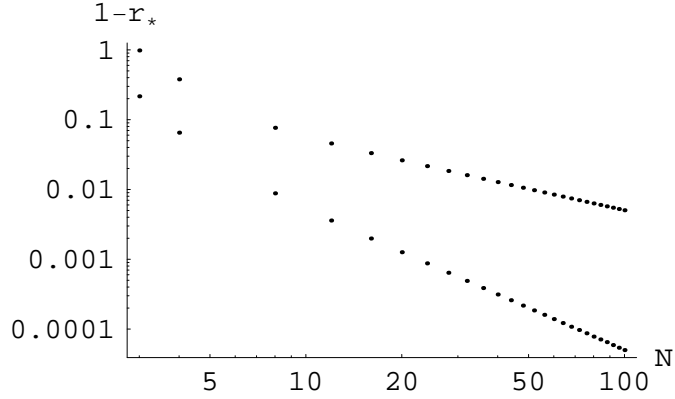}
  \caption{The logarithmic plot reports the behaviour of $1-r_*(N,N+1)$ 
    and $1-r_*(N,M_*(N))$ in the phase covariant broadcasting, for
    $3\leq N\leq100$. The upper line, corresponding to $M=M_*(N)$, has
    a power law $1/2N$.  The lower line, corresponding to $M=N+1$, has
    a power law $2/3N^2$.}\label{rstarphase}
\end{figure}

We will now discuss more quantitatively the results derived above.
The optimal scaling factors, reported in Eqs.~(\ref{eq:phase-pr1}) and
(\ref{eq:phase-pr2}), contain only known terms and can be studied
numerically.
\begin{figure}
  \includegraphics[width=9cm]{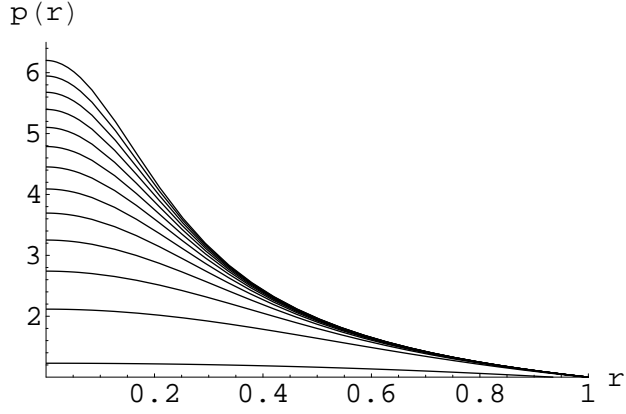}
  \caption{Optimal scaling factor $p^{N,N+1}(r)=r'/r$ versus $r$ for 
    the phase covariant broadcasting, for $M=N+1$ and $N$ ranging from
    4 to $100$ in steps of 8.}\label{pr-n+1}
\end{figure}
It turns out that phase covariant superbroadcasting is possible even
for $N=3$, with $M_*(3)=12$. Moreover, it is possible to
superbroadcast an infinite number of output copies starting from $N=4$
($M_*(N)=\infty$ for $N\ge 4$). As for the universal case, we can
easily compute the function $p(r)$ for $M=N+1$ and $M=\infty$, which
is monotone decreasing in $M$. In Fig. \ref{pr-n+1} we report the
plots of $p^{N,N+1}(r)$ for values of $N$ such that $4\le N\le 100$ in
steps of 8, and for $M=N+1$. In Fig.~\ref{rstarphase} we report the
plots of the values of $1-r_*(N,N+1)$ and $1-r_*(N,M_*(N))$, as
defined in the universal case.
The upper line refers to the case $N\to\infty$ and shows a behaviour
like $1/2N$. The lower line is for $N\to N+1$ and scales like
$2/(3N^2)$.

As before, in the limit of pure input states ($r=1$), the optimal
phase covariant superbroadcasting map coincides with the optimal phase
covariant cloner for qubits of Ref.~\cite{purequbitqutrit}.




\section{bipartite entanglement in the global output state}\label{sec:concurrence}

In this section we analyze the entanglement properties of the output
state of the optimal $N\to M$ broadcasting maps.  Notice that, since
broadcasting maps are always optimized with $j_l=M/2$, the output
state is supported on $\sH_{M/2}$ (which has multiplicity
$d_{M/2}=1$), namely the completely symmetric subspace of
$(\mathbb{C}^2)^{\otimes M}$. Therefore, also the reduced state of two
qubits $\rho^{(2)}=\Tr_{M-2}[\Sigma]$ is symmetric.  We will analyze
in particular bipartite entanglement in the output state, which is
conveniently described in terms of the concurrence \cite{wootters}
\begin{equation}
  C(\rho^{(2)})=\max\{0,\lambda_1-\lambda_2-\lambda_3-\lambda_4\}\,,
\end{equation}
where $\lambda_i$ are the decreasingly-ordered eigenvalues of the
operator $\Psi=\sqrt{\sqrt{\rho^{(2)}}\tilde\rho^{(2)}
  \sqrt{\rho^{(2)}}}$, and $\tilde\rho^{(2)}=\sigma_y^{\otimes
  2}\rho^{(2)}{}^*\sigma_y^{\otimes 2}$.

\begin{figure}[h]
\epsfig{file=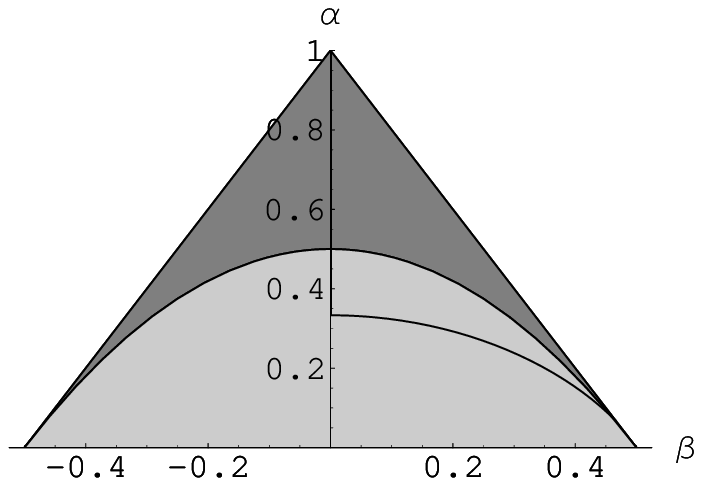,width=6.5cm}\epsfig{file=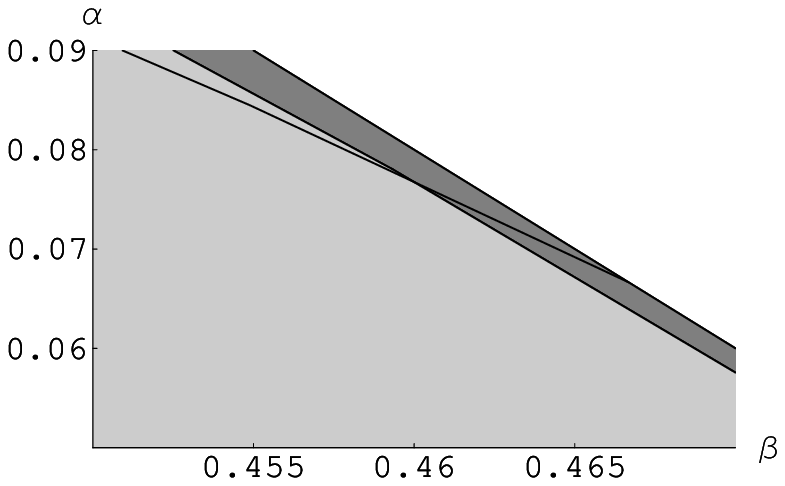,width=6.5cm}
\caption{\label{concurr} The figure represents the set of bipartite
  symmetric states which are diagonal on the $J_z^{(1)}$ basis. All
  such states are parametrized by the couple $(\beta,\alpha)$, as
  given in Eqs.~(\ref{eq:rho2}) and (\ref{eq:rho22}). The light gray
  region contains separable states. The dark gray region contains the
  entangled states. The black line is the parametric plot of the
  double-site reduced output of the $4\to 5$ universal broadcasting,
  for input Bloch vector length $r$ ranging from 0 to 1. In the
  magnified window on the right, it is possible to notice that, for
  nearly pure input, the black line crosses the parabola, namely, the
  output exhibits bipartite entanglement.}
\end{figure}

We will first consider the universal case, where the output state
$\Sigma$ is diagonal on the $J_z$ basis. As shown in
Appendix~\ref{appendixa}, the state $\rho^{(2)}$ commutes with
$J_z^{(1)}$, and therefore it can be written as a linear combination
of independent powers of $J^{(1)}_z$, namely
\begin{equation}\label{eq:rho2}
  \rho^{(2)}=\alpha \openone+\beta J_z^{(1)}+\gamma\left(J_z^{(1)}\right)^2\;.
\end{equation}
In the above expression the positivity and unit trace constraints are
given by
\begin{equation}\label{eq:rho22}
\alpha+\gamma\geq|\beta|,\quad\alpha\geq0,\quad3\alpha+2\gamma=1.
\end{equation}
The eigenvalue $\lambda_4$ of $\Psi$ is always 0, corresponding to the
null component of $\rho^{(2)}$ on the singlet. By the unit trace
condition we can express $\gamma$ as a function of $\alpha$ and $\beta$
as $\gamma=(1-3\alpha)/2$, and the positivity condition in terms
of the two independent parameters $(\beta,\alpha)$ is just
\begin{equation}\label{eq:triangle}
  \alpha\leq1-2|\beta|.
\end{equation}
The above inequality defines a triangle with basis $[-1/2,1/2]$ and
height $[0,1]$, as shown in Fig.~\ref{concurr} (left). A state in
$\sH_1$ is then completely determined by the couple $(\beta,\alpha)$.
Notice that the only pure states of the form (\ref{eq:rho2}) are
$\ket{1,1}$, $\ket{1,0}$ and $\ket{1,-1}$, which correspond to the
vertices $(1/2,0)$, $(0,1)$ and $(-1/2,0)$ respectively of the
triangle in Fig.~\ref{concurr}.

We will now express the concurrence in terms of $(\beta,\alpha)$.
Since $\rho^{(2)}$ is real it follows that
$\tilde\rho^{(2)}=\sigma_y^{\otimes 2}\rho^{(2)}\sigma_y^{\otimes 2}$,
and therefore we can write
\begin{equation}
  \tilde\rho^{(2)}=e^{i\pi J_y^{(1)}}\rho^{(2)} e^{-i\pi J_y^{(1)}}.
\end{equation}
It is easy to verify from Eq.~(\ref{eq:rho2}) that $\tilde\rho^{(2)}$
corresponds to the couple $(\alpha,-\beta)$. Moreover, since
$\rho^{(2)}$ and $\tilde\rho^{(2)}$ commute, the operator $\Psi$ can
be simply written as $\sqrt{\rho^{(2)}\tilde\rho^{(2)}}$.  By
exploiting some algebra, and taking into account the identities
$\left(J_z^{(1)}\right)^3=J_z^{(1)}$ and
$\left(J_z^{(1)}\right)^4=\left(J_z^{(1)}\right)^2$, we get the
following expression
\begin{equation}
\rho^{(2)}\tilde\rho^{(2)}=\alpha^2\openone
+(2\alpha\gamma+\gamma^2-\beta^2)\left(J_z^{(1)}\right)^2\;.
\end{equation}

>From the above expression, by using the unit trace constraint, we can
compute the eigenvalues of $\Psi$
\begin{equation}
  \left\{\frac{\sqrt{1-2\alpha+\alpha^2-4\beta^2}}2,\alpha,0\right\}.
\label{eigenv}
\end{equation}
Notice that the first eigenvalue is doubly degenerate. The concurrence
can then be written as follows
\begin{equation}
C(\rho)=\left\{
\begin{split}
&0\,,&0\leq\alpha\leq\frac{1-4\beta^2}2,\\
&\alpha-\sqrt{1-2\alpha+\alpha^2-4\beta^2}\,,&\alpha>\frac{1-4\beta^2}2.\\
\end{split}\right.
\end{equation}
The above equation defines a parabola inside the triangle
(\ref{eq:triangle}) of states. Such a parabola separates the region of
separable states from that of entangled states, shown in light and
dark gray in Fig.~\ref{concurr} respectively. In order to analyze the
amount of bipartite entanglement in the broadcast states, we have then
to evaluate the couple $(\beta,\alpha)$ for the reduced state of two
output copies and then determine in which region of the triangle it
lies. Using Eq. (\ref{alfabeto}) derived in Appendix C, we can
numerically evaluate $(\beta,\alpha)$ for the universal double-site
reduced output density operator $\rho^{(2)}=\Tr_{M-2}[\Sigma]$. In
Fig.~\ref{concurr} we report the parametric plot for the case of 4
input and 5 output copies. As we can see, the black line moves towards
positive $\beta$ as the Bloch vector length $r$ of the input state
goes from 0 to 1. It is possible to see in the magnified plot on the
right that, as $r$ gets close to 1, i.~e. the input state gets pure,
the output exhibits bipartite entanglement, since it crosses the
parabola. In the limit of pure input states, these results agree with
the ones derived in Ref.~\cite{entanglement-pure}.

In the phase covariant case it is not possible to carry on the same
analysis, since, as we notice in Appendix \ref{appendixa}, the global
output state does not commute with $J_x$. However, using the partial
traces in Eqs.~(\ref{alfabeto}),~(\ref{alfabeto2}),
and~(\ref{alfabeto3}), it is still possible to evaluate the
concurrence numerically.
\begin{figure}[h]
\epsfig{file=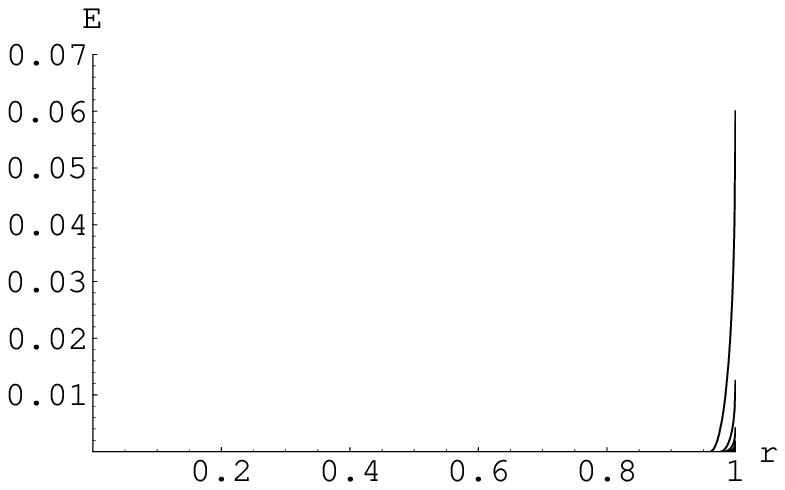,width=6.5cm}\epsfig{file=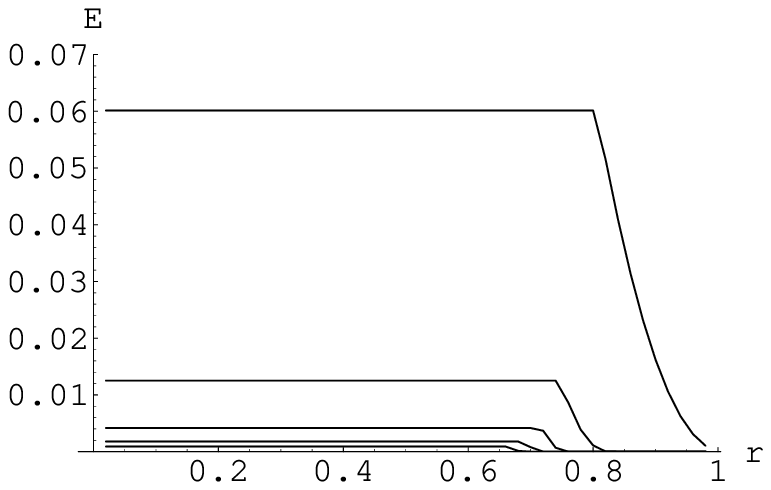,width=6.5cm}
\caption{\label{concuphase} Entanglement $E$ for the double-site
  reduced output state, in the cases of universal (l.h.s) and phase
  covariant (r.h.s) broadcasting from $N=2,4,6,8,10$ to $M=N+1$
  copies, as a function of the input Bloch vector length $r$.}
\end{figure}
In Fig.~\ref{concuphase} we report the plots of the entanglement $E$
defined in Ref. \cite{wootters} as follows
\begin{equation}
E=-\frac{1+\sqrt{1-C^2}}2\log\frac{1+\sqrt{1-C^2}}2-\frac{1-\sqrt{1-C^2}}2\log\frac{1-\sqrt{1-C^2}}2.
\end{equation}
for $N=2,4,6,8,10$ and $M=N+1$ as a function of the input Bloch vector
length $r$, both in the universal and phase covariant case.  Notice
that, contrarily to what happens in the universal case, in the phase
covariant case bipartite entanglement vanishes in the limit of pure
input states. The absolute value of $C$ goes to zero for increasing
number $N$ of input copies.

\section{Conclusions and further developments}\label{s:conc}

In this paper we studied symmetric broadcasting maps, where $N$ input
qubits initially prepared in the same mixed state are transformed into
an output state of $M$ qubits, all described by the same density
operator.  We considered covariant maps and we investigated in
particular the universally covariant case and the phase covariant
case. We have shown that for sufficiently mixed initial states and for
$N\geq 3$ it is also possible to partially purify the single qubit
output density operator in the broadcasting operation. Such a new
process was named superbroadcasting.

The new superbroadcasting channels open numerous interesting
theoretical problems. The first problem is to extend the map to any
dimension $d>2$ and to different covariance groups. Indeed, for
special cases it is easy to see that increasing the dimensionality
and/or reducing the set of states to be broadcast makes
superbroadcasting possible with smaller $N$, and even with $N=2$ input
states. As a matter of fact, this is the case of universally covariant
superbroadcasting from $N=4$ to $M=6$, which can also be regarded as
superbroadcasting for $d=4$ for special states of the form
$\rho\otimes\rho$, and for the covariance group
$\mathbb{SU}(2)\times\mathbb{SU}(2)$. The case of dimension $d=4$ is
most interesting, since it can be exploited to improve entanglement
for bipartite states of qubits. Also the infinite dimensional case
(the so-called ``continuous variables'') turns out to be interesting,
and easily feasible experimentally \cite{cvsb}. It should be
emphasized that for dimension $d\ge3$ there are many ways of
increasing purity, and certainly the most interesting case is the
purification along the mixing direction of a noisy channel (notice
that most channels do not correlate, whence the produced state is the
tensor product of identical mixed states).

Another major problem is the analysis of the detrimental correlations
between two outputs, e. g. to establish whether they are quantum or classical.
These correlations are exotic, in the sense that instead of increasing
the local mixing as usual, they reduce it. Such mechanism is new, and
deserves a more thorough analysis. An interesting issue, for example,
is that they cannot be erased leaving the local state unchanged (the
de-correlating map---which sends a state to the tensor product of its
partial traces---is non linear), and this raises the problem of the
optimal de-correlating channel, which optimizes the fidelity between
the input and the output local state. Such optimal channel can be
derived using the same technique for optimal covariant maps used in
the present paper.

Finally, distributing quantum information---and in particular
superbroadcasting---raises the new problem of the trade-off between
broadcasting and cryptographic security. Indeed, on one side, the
presence of many identical uses seems to open more possibilities of
eavesdropping, however the detrimental correlations may drastically
reduce such possibility, and the opportunity of detecting the
eavesdropping on the joint output state may be exploited to increase
the security.  

\acknowledgments This work has been co-founded by the EC under the
program SECOQC (Contract No. IST-2003-506813) and by the Italian MIUR
through FIRB (bando 2001) and PRIN 2005.

\appendix
\section{Decomposition of $\rho^{\otimes N}$}\label{appendixaa}
The global density operator $\rho^{\otimes N}$ is clearly invariant
under permutations of the $N$ qubits, and, according to the Schur-Weyl
duality, it can be represented by the Wedderburn decomposition
(\ref{eq:schur-Weyl-dual-form})
\begin{equation}
\rho^{\otimes N}=\bigoplus_{j=j_0}^{N/2}\rho_j\otimes \openone_{d_j},
\end{equation}
where $\rho_j$ is a state on $\sH_j$. In order to evaluate $\rho_j$ it
is sufficient to evaluate the matrix elements
\mbox{$\<jm|\otimes\<\alpha|\rho^{\otimes
    N}|jm^\prime\>\otimes|\alpha\>$}, where $|jm\>$ are eigenstates of
$J_z$ in the $j$ representation, and $|\alpha\>$ is an arbitrary state
in $\mathbb{C}^{d_j}$.  For the sake of simplicity we can suppose that
the state $\rho$ has the Bloch form $\frac12(\openone+r\sigma_z)$. The
problem is now to choose $|\alpha\>$ in a suitable way. It turns out
that a clever choice is given by
\begin{equation}\label{eq:clever-choice}
  |jm\>\otimes|\alpha\>=|jm\>_{2j}\otimes|\Psi^-\>^{\otimes \frac{N-2j}2},
\end{equation}
where the subscript $2j$ means that $|jm\>_{2j}$ is a vector in the
symmetric subspace of the first $2j$ qubits, while $|\Psi^-\>$ is a
singlet, supporting an invariant representation on a couple of qubit
spaces. Notice that in Eq.~\eqref{eq:clever-choice} the tensor product
on the l.h.s. refers to the ``abstract'' subspace
$\sH_j\otimes\mathbb{C}^{d_j}$ in the Wedderburn decomposition,
whereas the one on the r.h.s. refers to the decomposition
$(\mathbb{C}^{2})^{\otimes 2j}\otimes(\mathbb{C}^2)^{\otimes(N-2j)}$
grouping separately the first $2j$ qubits and the remaining $N-2j$. Moreover,
since the chosen density operator $\rho^{\otimes N}$ commutes with the total 
$J_z$, given by the following expression
\begin{equation}\label{eq:tot-ang-momentum}
J_z=\bigoplus_{j=j_0}^{N/2}J_z^{(j)}\otimes \openone_{d_j}=\bigoplus_{j=j_0}^{N/2}\sum_{m=-j}^jm|jm\>\<jm|\otimes \openone_{d_j},
\end{equation}
then $\rho_j$ commutes with $J_z^{(j)}$, which implies that $\rho_j$
is diagonal on the eigenstates $|jm\>$ of $J_z^{(j)}$. Therefore we can write
\begin{equation}
  (\rho_j)_{mm}=\;{}_{2j}\<jm|\rho^{\otimes 2j}|jm\>_{2j}\ \left(\<\Psi^-\left|\rho^{\otimes 2}\right|\Psi^-\>\right)^{\frac{N-2j}2}\,.
\end{equation}
Since $|jm\>_{2j}$ is symmetric, it is a linear combination of
factorized vectors with $(j+m)$ qubits in the $|1/2,1/2\>$ state and
$(j-m)$ in the state $|1/2,-1/2\>$. As a consequence we can also write
\begin{equation}
\rho^{\otimes 2j}|jm\>_{2j}=r_+^{j+m}r_-^{j-m}|jm\>_{2j}\,,
\end{equation}
where $r_\pm=\frac{1\pm r}2$. By analogous arguments it follows that
\begin{equation}
\rho^{\otimes 2}|\Psi^-\>=r_+r_-|\Psi^-\>\,.
\end{equation}
The matrix element $(\rho_j)_{mm}$ has the following expression
\begin{equation}\label{rearr}
  (\rho_j)_{mm}=r_+^{j+m}r_-^{j-m}(r_+r_-)^{N/2-j}=(r_+r_-)^{N/2}\left(\frac{r_+}{r_-}\right)^m,
\end{equation}
and the decomposition of $\rho^{\otimes N}$ is finally given by
\cite{Cirac99}
\begin{equation}
  \rho^{\otimes N}=(r_+r_-)^{N/2}\bigoplus_{j=j_0}^{N/2}\sum_{m=-j}^j\left(\frac{r_+}{r_-}\right)^m|jm\>\<jm|\otimes \openone_{d_j}\,.
\end{equation}
Notice that this expression exhibits a singularity for $r=1$ due to
the rearrangement of terms (\ref{rearr}). However, a finite limit for
$r\to 1$ exists, as it can be seen from the equivalent expression
\begin{equation}\label{eq:rho-decomposition2}
\rho^{\otimes N}=\bigoplus_{j=j_0}^{N/2}(r_+r_-)^{N/2-j}\sum_{m=-j}^jr_+^{j+m}r_-^{j-m}|jm\>\<jm|\otimes\openone_{d_j}\,,
\end{equation}
which exhibits no singularities.

\section{Formulae for the scaling factors}\label{appendixb}
In this Appendix we will derive the explicit form of the scaling factor for
the universal and phase covariant cases.
\subsection{Universal case}
In order to calculate $p^{N,M}(r)$ we start rewriting Eq.
(\ref{eq:to-calculate}) as
\begin{equation}
  \Tr\left[\left(J_z^{(j)}\otimes|ln\>\<ln|\right)P^{J}_{j,l}\right]=\Tr\left[|ln\>\<ln|\Tr_j\left[\left(J_z^{(j)}\otimes\openone_{2l+1}\right)\ P^{J}_{j,l}\right]\right].
\end{equation}
Let us define
\begin{equation}
X_i^{(l)}\equiv\Tr_{j}\left[\left(J_i^{(j)}\otimes\openone_{2l+1}\right)\ P^{J}_{j,l}\right],
\end{equation}
where $i=-1,0,1$, and $X_0\doteq X_z$. Since
\begin{equation}
\left(U_g^{(j)}\otimes U^{(l)}_g\right)P^{J}_{j,l}\left(U_g^{(j)}\otimes U^{(l)}_g\right)^\dag=P^{J}_{j,l},
\end{equation}
the set $X_i^{(l)}$ transforms according to
\begin{equation}
\begin{split}
  U^{(l)}_g X_i^{(l)}
  {U_g^{(l)}}^\dag&=\Tr_{j}\left[\left(U_g^{(j)}J_i^{(j)}{U_g^{(j)}}^\dag\otimes\openone_{2l+1}\right)\ P^{J}_{j,l}\right]\\
&=\Tr_{j}\left[\left(\sum_{k=-1}^1(U^{(1)}_g)_{ik}J_k^{(j)}\otimes\openone_{2l+1}\right)\
  P^{J}_{j,l}\right]\\
  &=\sum_{k=-1}^1(U^{(1)}_g)_{ik}X_k^{(l)},\\
\end{split}
\end{equation}
and we conclude that $\left\{X_i^{(l)}\right\}$ is an irreducible
tensor set. It can then be proved by the Wigner-Eckart theorem that
$X_i^{(l)}=\alpha J_i^{(l)}$, and in particular $X_z^{(l)}=\alpha
J_z^{(l)}$. From the last relation and from the identity
\begin{equation}
  \frac{1}{2}\left(J_+^{(l)}J_-^{(l)}+J_-^{(l)}J_+^{(l)}\right)+\left(J_z^{(l)}\right)^2=\sum_{i=-1}^1a_iJ_i^{(l)}J^{(l)}_{-i}=\left(J^{(l)}\right)^2\equiv
  l(l+1)\openone_{2l+1},
\end{equation}
where $a_{-1}=a_1=1/2$ and $a_0=a_z=1$, and $J^{(k)}_0\doteq
J^{(k)}_z$, we have
\begin{equation}
\begin{split}
\alpha l(l+1)(2l+1)&=\alpha\sum_{i=-1}^1a_i\Tr\left[J_i^{(l)}J_{-i}^{(l)}\right]\\
&=\sum_{i=-1}^1a_i\Tr\left[\left(J_i^{(j)}\otimes J_{-i}^{(l)}\right)P^J_{j,l}\right],
\end{split}
\end{equation}
By using the well known identity
\begin{equation}
\begin{split}
  \sum_{i=-1}^1a_i\left(J_i^{(j)}\otimes J^{(l)}_{-i}\right)\
  P^{J}_{j,l}&=\frac12\left({J^{(J)}}^2-{J^{(j)}}^2\otimes\openone_{2l+1}-\openone_{2j+1}\otimes{J^{(l)}}^2\right)P^{J}_{j,l}\\
  &=P^{J}_{j,l}\frac{J(J+1)-j(j+1)-l(l+1)}{2}\,,
\end{split}
\end{equation}
we can write the explicit form of the coefficient $\alpha$
\begin{equation}
\alpha=\frac{2J+1}{2l+1}\frac{J(J+1)-j(j+1)-l(l+1)}{2l(l+1)}\equiv\frac{2J+1}{2l+1}\beta(J,j,l).
\end{equation}
By using the above expression we finally have
\begin{equation}
  \Tr\left[\left(J_z^{(j)}\otimes|ln\>\<ln|\right)\ P^{J}_{j,l}\right]=n\frac{2J+1}{2l+1}\beta(J,j,l).
\end{equation}

\subsection{Phase covariant case}
Substituting the global output state $\Sigma$
given by Eq. (\ref{eq:form-for-Sigma}) into Eq.~(\ref{eq:rprime-phase}), it is
possible to compute the scaling factor $p^{N,M}(r)$ as
\begin{equation}
\begin{split}
  p^{N,M}(r)\equiv\frac{r'}{r}=&\frac 1r\frac{2}{M}(r_+r_-)^{N/2}\sum_{j,l,k}d_jd_l\sum_{n,n'=-l}^lr^{j,l}_{n,n',k}\left[J_x^{(j)}\right]_{n'+k,n+k}\times\\
&\sum_{n''=-l}^l(W_l)_{n',n''}\left(\frac{r_+}{r_-}\right)^{n''}(W_l^\dag)_{n'',n},
\end{split}
\label{app1}
\end{equation}
where $(W_k)_{ab}\equiv\<k,a|k^x,b\>$ are entries of the Wigner
rotation matrix in the $k$ representation which rotates the $z$-components
to the $x$-components (in the usual notation, the one that is found, for
example, in \cite{edmonds}, such entries are denoted as
$d^{(k)}_{ab}\left(\beta\equiv\frac{\pi}{2}\right)$). 
In Eq. (\ref{app1}) the sum over $n''$ gives the $(n,n')$-th matrix element of
$\exp(J_x^{(l)}\log r_+/r_-)$. Therefore we can write
\begin{equation}\label{eq:phase-form-pr}
\begin{split}
  p^{N,M}(r)&=\frac{2}{Mr}(r_+r_-)^{N/2}\sum_{j,l,k}d_jd_l\sum_{n,n'=-l}^lr^{j,l}_{n,n',k}\left[\exp\left(J_x^{(l)}\log\frac{1+r}{1-r}\right)\right]_{n',n}\left[J_x^{(j)}\right]_{n'+k,n+k}\\
  &=\frac{4}{Mr}(r_+r_-)^{N/2}\sum_{j,l,k}d_jd_l\sum_{n=-l}^lr^{j,l}_{n,n+1,k}\left[\exp\left(J_x^{(l)}\log\frac{1+r}{1-r}\right)\right]_{n,n+1}\left[J_x^{(j)}\right]_{n+k,n+k+1},
\end{split}
\end{equation}
where in the last equality we used the fact that
$J_x^{(j)}=\sum_mm|j^x,m\>\<j^x,m|$ has non-null matrix elements only
on the second-diagonals, and we multiplied the second line by a factor 2
considering in the sum only one of the two second-diagonals. Notice
that all matrix elements in the previous equations are calculated with
respect to the $z$-oriented basis $|j^z,m\>\equiv|j,m\>$ and
$|l^z,n\>\equiv|l,n\>$.

\section{Reduced output states}\label{appendixa}
\subsection{Single-site reduced output states}
In this appendix we want to calculate the following partial trace
\begin{equation}
\Tr_{M-1}\left[|jm\>\<jm|\otimes\openone_{d_j}\right],
\end{equation}
where the operator to be partially traced acts on
$\sH_j\otimes\mathbb{C}^{d_j}\subset(\mathbb{C}^2)^{\otimes M}$, and
$|jm\>$ are eigenstates of $J_z^{(j)}$, as usual. In order to do that,
we first decompose the vector $|jm\>$ into its components onto
$\sH_{j-1/2}\otimes\sH_{1/2}$ using the Clebsch-Gordan coefficients
\begin{equation}\label{eq:cg-coeff}
  |jm\>=\sqrt{\frac{j+m}{2j}}\left|j-\frac 12,m-\frac 12\right\rangle\otimes\left|\frac 12,\frac 12\right\rangle+\sqrt{\frac{j-m}{2j}}\left|j-\frac 12,m+\frac 12\right\rangle\otimes\left|\frac 12,-\frac 12\right\rangle,
\end{equation}
and then trace the operator $|jm\>\<jm|$ over $\sH_{j-1/2}$. In this
way we get
\begin{equation}\label{partial-trace}
  \Tr_{j-1/2}\left[|jm\>\<jm|\right]=\frac{\openone}{2}+\frac{m}{2j}\sigma_z.
\end{equation}

We now recall a fact related to the already mentioned Schur-Weyl
duality, by which multiplicity spaces $\mathbb{C}^{d_j}$ in the
Wedderburn decomposition (\ref{eq:wedderburn-for-su2}) support
irreducible representations of the permutation group
$\{\Pi_\sigma^M\}$ of $M$ qubits. Hence, for any operator $O$ on
$\sH_j\otimes\mathbb{C}^{d_j}$ one has
\begin{equation}
  \sum_\sigma\Pi_\sigma^M O\Pi_\sigma^M=\frac{M!}{d_j}\Tr_{\mathbb{C}^{d_j}}[O]\otimes\openone_{d_j}.
\end{equation}
For convenience, let us write
\begin{equation}
|jm\>\<jm|\otimes\openone_{d_j}=\frac{d_j}{M!}\sum_\sigma\Pi_\sigma^M\left(|jm\>\<jm|\otimes|\Psi^-\>\<\Psi^-|^{\otimes\frac{M-2j}{2}}\right)\Pi_\sigma^M,
\end{equation}
as we already did in Eq.~(\ref{eq:clever-choice}). With this choice,
we get:
\begin{equation}\label{eq:very-similar}
\begin{split}
\Tr_{M-1}\left[|j,m\>\<j,m|\otimes\openone_{d_j}\right]&=\Tr_{M-1}\left[\frac{d_j}{M!}\sum_\sigma\Pi_\sigma^M\left(|j,m\>\<j,m|\otimes|\Psi^-\>\<\Psi^-|^{\otimes\frac{M-2j}{2}}\right)\Pi_\sigma^M\right]\\
&=\frac{d_j}{M!}\left[(M-1)!(M-2j)\frac{\openone}{2}+(M-1)!2j\left(\frac{\openone}{2}+\frac{m}{2j}\sigma_z\right)\right]\\
\end{split}
\end{equation}
The first term in the sum comes from excluding from trace one of the
$(M-2j)$ qubits in singlet state. The second term in the sum comes
from excluding from trace one of the $2j$ qubits in $|jm\>$ state, and
from Eq.~(\ref{partial-trace}). Rearranging the above equation, we get
the final expression
\begin{equation}\label{eq:final-partial-trace}
\Tr_{M-1}\left[|j,m\>\<j,m|\otimes\openone_{d_j}\right]=d_j\left(\frac{\openone}{2}+\frac{m}{M}\sigma_z\right).
\end{equation}
\subsection{Properties of single-site output states}
Consider the global output state $\Sigma=\mathcal{B}(\rho^{\otimes
  N})$. If it commutes with the total angular momentum component along
the direction $z$, for example, then it is simple to prove that also
$\rho'=\Tr_{M-1}[\Sigma]$ commutes with $\sigma_z$. In fact, from
Eq.~(\ref{eq:final-partial-trace}), it is simple to see that
\begin{equation}
  \Tr_{M-1}\left[J_z^{(j)}\otimes\openone_{d_j}\right]\propto\sigma_z,
\end{equation}
for all $j$, and consequently also
$\Tr_{M-1}\left[J_z\right]\propto\sigma_z$.

In the universal case, the global output $\Sigma$ is diagonal on
eigenstates of $J_z^{(M/2)}$, hence $\rho'=\Tr_{M-1}[\Sigma]$ commutes
with $\sigma_z$, according to previous arguments. In the phase
covariant case, it is more difficult to prove on general grounds that
$[\rho',\sigma_x]=0$, since in the phase covariant case
$\left[J_x,\Sigma\right]\neq 0$. The simplest thing we can do is to
compute the partial trace of Eqs.~(\ref{eq:phase-pr1}) and
(\ref{eq:phase-pr2}) using again Clebsch-Gordan coefficients
(\ref{eq:cg-coeff}). First of all let us notice that, tracing over
$M-1$ qubits, only the terms with $|n-n'|\le 1$ contribute. Among
these, the terms with $n=n'$ give one factor proportional to
$\openone/2$ and one factor proportional to $\sigma_z$, whereas terms
with $|n-n'|=1$ contribute with factors proportional to $\sigma_x$,
since the matrix of coefficients
$\exp\left(J_x^{(l)}\log\frac{1+r}{1-r}\right)$ is symmetric, see
Eq.~(\ref{eq_che_manca}). Then, posing $\alpha_k=0$ without loss of
optimality, $\Tr_{M-1}[\Sigma]$ commutes with $\sigma_x$.

\subsection{Double-site reduced output states}
We will show here how to compute the reduced output state of two
copies, which is used in Section \ref{sec:concurrence} to compute the 
concurrence between two of the $M$ clones. Clearly it does not matter which 
two clones we are
considering, since the global output state is permutation invariant.
Using Clebsch-Gordan coefficients, it is possible to decompose a
vector in $\sH_j$ into its components onto $\sH_{j-1}\otimes\sH_1$
\begin{equation}
\begin{split}
|jm\>=&\sqrt{\frac{(j+m)(j+m-1)}{2j(2j-1)}}|j-1,m-1\>|1,1\>+\sqrt{\frac{j^2-m^2}{j(2j-1)}}|j-1,m\>|1,0\>+\\&\sqrt{\frac{(j-m)(j-m-1)}{2j(2j-1)}}|j-1,m+1\>|1,-1\>,
\end{split}
\end{equation}
and then to compute the partial trace of $|jm\>\<jm'|$ over
$\sH_{j-1}$. For $j=M/2$ we have
\begin{equation}\label{alfabeto}
\Tr_{M-2}\left[|M/2,m\>\<M/2,m|\right]=\begin{pmatrix}
\frac{(M-2m)(M-2m-2)}{4M(M-1)}&0&0\\
0&\frac{M^2-4m^2}{2M(M-1)}&0\\
0&0&\frac{(M+2m)(M+2m-2)}{4M(M-1)}
\end{pmatrix},
\end{equation}
when $|m-m'|=1$
\begin{equation}\label{alfabeto2}
\begin{split}
\Tr_{M-2}&\left[|M/2,m\>\<M/2,m+1|+\textrm{h.c.}\right]=\\
&\begin{pmatrix}
0&\frac{(M-2m-2)\sqrt{(M-2m)(M+2m+2)}}{2\sqrt{2}M(M-1)}&0\\
\frac{(M-2m-2)\sqrt{(M-2m)(M+2m+2)}}{2\sqrt{2}M(M-1)}&0&\frac{(M+2m)\sqrt{(M-2m)(M+2m+2)}}{2\sqrt{2}M(M-1)}\\
0&\frac{(M+2m)\sqrt{(M-2m)(M+2m+2)}}{2\sqrt{2}M(M-1)}&0
\end{pmatrix},
\end{split}
\end{equation}
and $|m-m'|=2$,
\begin{equation}\label{alfabeto3}
\begin{split}
\Tr_{M-2}&\left[|M/2,m\>\<M/2,m+2|+\textrm{h.c.}\right]=\\
&\begin{pmatrix}
0&0&\frac{\sqrt{(M-2m)(M+2m+4)(M+2m+2)(M-2m-2)}}{4M(M-1)}\\
0&0&0\\
\frac{\sqrt{(M-2m)(M+2m+4)(M+2m+2)(M-2m-2)}}{4M(M-1)}&0&0
\end{pmatrix}.
\end{split}
\end{equation}
For $|m-m'|\ge 3$, partial trace over $M-2$ copies gives null
contribution.


\begin{thebibliography}{99}
\bibitem{discom} L. Grover, quant-ph/9704012; J I Cirac, A Ekert, S F
  Huelga, and C. Macchiavello, Phys. Rev. A {\bf 59}, 4249 (1999).
\bibitem{shacrip} C Crepeau, D Gottesman, and A Smith, in Proceedings
  of the 34th Annual ACM Symposium on Theory of Computing (New York,
  ACM Press, 2002).
\bibitem{gam} S C Benjamin and P M Hayden, Phys. Rev. A {\bf 64},
  030301 (2001).
\bibitem{Wootters82} W K Wootters, W H Zurek, Nature {\bf 299}, 802
  (1982).
\bibitem{Dieks82}  D Dieks, Phys. Lett. A, {\bf 92}, 271 (1982).
\bibitem{Yuen} H P Yuen, Phys.\ Lett.\ A{\bf 113} 405 (1986).
\bibitem{noteclon} In Ref.~\cite{Wootters82} it was shown that the
  cloning machine violates the superposition principle, which applies
  to a minimum total number of {\em three} states, and hence does not
  rule out the possibility of cloning {\em two} nonorthogonal states.
  It is violation of unitarity that makes cloning any {\em two}
  nonorthogonal states impossible, as proved in Ref.~\cite{Yuen}.  In
  reference \cite{Dieks82} it was shown that the proposal for
  superluminal communication does not work, proving the impossibility
  of cloning in this particular context due to linearity of evolution.
  According to Ref.~\cite{peres}, previous to Refs.
  \cite{Wootters82,Dieks82} the anonymous referee's report of G.
  Ghirardi to Ref.~\cite{herbert} contained an argument which was a
  special case of the no-cloning theorem of Refs.
  \cite{Wootters82,Dieks82}. More recently, after several attempts of
  determining the wave function of a single system appeared in the
  literature, Ref.~\cite{darianoyuen} showed how it is impossible to
  determine the wave function from a single copy of the system, and
  connected such impossibility to the no-cloning theorem.
\bibitem{darianoyuen} G M D'Ariano and H P Yuen, Phys. Rev. Lett.
  {\bf 76} 2832 (1996).
\bibitem{herbert} N Herbert, Found. Phys. {\bf 12}, 1171 (1982).
\bibitem{peres} A Peres, Fortsch. Phys. {\bf 51}, 458 (2003).
\bibitem{Barnum96} H Barnum, C M Caves, C A Fuchs, R Jozsa, and B
  Schumacher, Phys. Rev. Lett. {\bf 76} 2818 (1996).
\bibitem{clifton} R Clifton, J Bub, and H Halvorson, Found. of Phys.
  {\bf 33} 1561 (2003).
\bibitem{our} G M D'Ariano, C Macchiavello, and P Perinotti, Phys.
  Rev. Lett. {\bf 95}, 060503 (2005).
\bibitem{Werner98} R F Werner, Phys. Rev. A {\bf 58}, 1827 (1998).
\bibitem{Keyl01} M. Keyl and R. F. Werner, Ann. H. Poincar\'e, {\bf
    2}, 1 (2001).
\bibitem{cvsb} G M D'Ariano, P Perinotti, and M F Sacchi,
  quant-ph/0602037; G M D'Ariano, P Perinotti, and M F Sacchi,
  quant-ph/0601114.
\bibitem{Cirac99} J I Cirac, A K Ekert, and C Macchiavello, Phys.
  Rev. Lett. {\bf 82}, 4344-4347 (1999).
\bibitem{altro} F Buscemi, G M D'Ariano, C Macchiavello, and
  P~Perinotti, in Proceedings of the 13th Quantum Information
  Technology Symposium (QIT13) (Sendai, Japan, 2005).
\bibitem{Buscemi03} F Buscemi, G M D'Ariano, and M F Sacchi, Phys.
  Rev. A {\bf 68}, 042113 (2003).
\bibitem{choi-jam} A Jamio\l kowski, Rep. Math. Phys. {\bf 3}, 275
  (1972); M-D Choi, Lin. Alg. Appl. {\bf 10}, 285 (1975).
\bibitem{fulton} W Fulton and J Harris, \emph{Representation Theory: a
    First Course} (Springer-Verlag, Berlin, 1991).
\bibitem{edmonds} A R Edmonds, \emph{Angular Momentum in Quantum
    Mechanics} (Princeton University Press, Princeton, 1960).
\bibitem{unot} V Bu\v{z}ek, M Hillery, and R F Werner, Phys. Rev. A
  {\bf 60}, R2626 (1999).
\bibitem{purequbitqutrit} G M D'Ariano and C Macchiavello, Phys. Rev.
  A {\bf 67}, 042306 (2003).
\bibitem{phasenot} F Buscemi, G M D'Ariano, and C Macchiavello, Phys.
  Rev A {\bf 72}, 062311 (2005).
\bibitem{wootters} W K Wootters, Phys. Rev. Lett. {\bf 80}, 2245
  (1998).
\bibitem{entanglement-pure} D Bru\ss{} and C Macchiavello, Found.
  Phys. {\bf 33} (11), 1617 (2003).
\end{thebibliography}
\end{document}